\documentclass[lettersize,journal]{IEEEtran}
\usepackage{amsmath, graphics,amssymb,epsfig, subfigure}
\usepackage{amsfonts,amssymb}
\usepackage{graphicx, color, rotating}
\usepackage{multirow}
\usepackage{bm}      
\usepackage{hhline}
\usepackage{cite}
\usepackage[table,x11names]{xcolor}  
\usepackage{tikz}                    

\usepackage{algorithm}
\usepackage{algorithmic}

\usepackage{cases}
\usepackage{booktabs}
\usepackage{multirow}
\usepackage{adjustbox}
\usepackage{nicematrix}
\usepackage{textcomp}
\usepackage{soul}  

\newcommand{\muldot}{%
    \tikz[baseline=(multiply.base)]{%
        \node[inner sep=0pt] (multiply) {$\cdot$};
        \draw (multiply) circle (3pt);
    }%
}

\newcommand*{\mydprime}{^{\prime\prime}\mkern-1.2mu}

\makeatletter
\def\hlinewd#1{%
  \noalign{\ifnum0=`}\fi\hrule \@height #1 \futurelet \reserved@a\@xhline}

\makeatother

\begin{document}

\title{SNR-aware Semantic Image Transmission with Deep Learning-based Channel Estimation in Fading Channels}

\author{Mahmoud M. Salim, Mohamed S. Abdalzaher, \IEEEmembership{Senior Member, IEEE}, Ali H. Muqaibel, \IEEEmembership{Senior Member, IEEE}, Hussein A. Elsayed, and Inkyu Lee, \IEEEmembership{Fellow, IEEE}

\thanks{Mahmoud M. Salim is with the Center for Communications Systems and Sensing, King Fahd University of Petroleum and Minerals (KFUPM), Dhahran 31261, Saudi Arabia, and also with the Department of Electrical Engineering, October 6 University, Giza 12585, Egypt (e-mails: mahmoud.elemam@kfupm.edu.sa; m.salim.eng@o6u.edu.eg).
Ali H. Muqaibel is with the Center for Communications Systems and Sensing and the Department of Electrical Engineering, KFUPM, Dhahran 31261, Saudi Arabia (e-mail: muqaibel@kfupm.edu.sa).}
\thanks{Mohamed S. Abdalzaher is with the Seismology Department, National Research Institute of Astronomy and Geophysics, Cairo 11421, Egypt (E-mail: msabdalzaher@nriag.sci.eg).}
\thanks{Hussein A. Elsayed is with the Department of Electronics and Communications, Faculty of Engineering, Ain Shams University, Cairo, 11566, Egypt (E-mail: helsayed@eng.asu.edu.eg).}
\thanks{Inkyu Lee is with the School of Electrical Engineering,
Korea University, Seoul 02841, South Korea (e-mail: inkyu@korea.ac.kr). Inkyu Lee is the corresponding author.}}
\maketitle
\begin{abstract}
Semantic communications (SCs) play a central role in shaping the future of the sixth generation (6G) wireless systems, which leverage rapid advances in deep learning (DL). In this regard, end-to-end optimized DL-based joint source-channel coding (JSCC) has been adopted to achieve SCs, particularly in image transmission. Utilizing vision transformers in the encoder/decoder design has enabled significant advancements in image semantic extraction, surpassing traditional convolutional neural networks (CNNs). In this paper, we propose a new JSCC paradigm for image transmission, namely Swin semantic image transmission (SwinSIT), based on the Swin transformer. The Swin transformer is employed to construct both the semantic encoder and decoder for efficient image semantic extraction and reconstruction. Inspired by the squeezing-and-excitation (SE) network, we introduce a signal-to-noise-ratio (SNR)-aware module that utilizes SNR feedback to adaptively perform a double-phase enhancement for the encoder-extracted semantic map and its noisy version at the decoder. Additionally, a CNN-based channel estimator and compensator (CEAC) module repurposes an image-denoising CNN to mitigate fading channel effects. To optimize deployment in resource-constrained IoT devices, a joint pruning and quantization scheme compresses the SwinSIT model. Simulations evaluate the SwinSIT performance against conventional benchmarks demonstrating its effectiveness. Moreover, the model's compressed version substantially reduces its size while maintaining favorable PSNR performance.
\end{abstract}

\begin{IEEEkeywords}
Semantic communications, joint source-channel coding, Swin transformer, image transmission.
\end{IEEEkeywords}

\section{Introduction} \label{Sec_Intro}

\IEEEPARstart{S}{emantic} communications (SC) is an innovative approach that attracts the attention of both the industry and academia to fulfill high requirements of the sixth generation (6G) networks \cite{9955525,chaccour-2024}. Virtual reality, augmented reality, holographic projection, high-definition image and video, and other high-capacity multimodal services are among the applications of 6G. This comes with the need for intelligent, collaborative, and personalized communications \cite{baek-2022-2,lu-2023scS,liang-2024SC,salim-2024}. 

Source coding and channel coding are the two main stages of conventional wireless communication systems \cite{yilmaz2022distributed}. 
For instance, most wireless image transmission relies on compression techniques such as JPEG for source coding, as well as turbo codes \cite{berrou-1996} or polar coding \cite{hussami2009performance} for channel coding. Joint source-channel coding (JSCC) has recently become a key factor in allowing graceful performance adaptation even in the presence of channel changes \cite{7976323,8723589}. Due to lower computational costs and shorter delays than separate source-channel coding, the latter is preferable in situations where computational effectiveness and real-time processing are crucial \cite{7086847,504941}. In the context of SC, JSCC makes use of intelligent deep learning (DL)-based architectures to extract significant semantics and transmit them through fading channels \cite{yang2022semantic, 10038657}. This makes tasks like reconstruction, classification, or segmentation easier to complete, with notable advantages in image transmission \cite{9913352, 9955525}.

The combination of computer vision and DL has allowed substantial improvements in the design of semantic encoders and decoders in recent years. Convolutional neural networks (CNNs) excel at extracting image semantics, impacting semantic perception \cite{9953099}, whereas autoencoders \cite{schmidhuber-2015} play a crucial role in compressing and reconstructing semantic information due to their ability to capture compact representations \cite{8723589}. Recent advances in semantic processing introduced attention-based vision transformers, such as the Swin transformer, which can handle high-resolution (HR) images with complex semantic content \cite{10094735,liu2021Swin}.

 In the domain of semantic communication for image transmission, accurate channel estimation is crucial to ensuring communication reliability by preserving the semantic content of transmitted images \cite{kong-2017}. Traditional estimation techniques, such as maximum likelihood (ML) and least squares, are widely used for their simplicity and computational efficiency. However, these methods often face limitations in practical scenarios, as they do not effectively account for noise in the channel estimation process \cite{SURE2017629}. To address these limitations, advanced methods like minimum mean square error (MMSE) and linear MMSE (LMMSE) are generally preferred because they incorporate noise statistics, enabling more accurate estimates in fading environments \cite{coleri2002channel}. A common practical approach involves obtaining initial channel estimates using ML or least squares estimators, which are subsequently refined through advanced techniques like MMSE or LMMSE for improved accuracy \cite{8640815}. Recently, DL has emerged as a promising tool for enhancing channel estimation by leveraging its ability to model complex relationships in noisy environments \cite{o2017introduction, 8640815, zeng2019toward}. Specifically, DL-based image denoising models can take channel parameters derived from ML or least squares estimators as inputs, effectively replacing conventional refinement methods such as MMSE. This approach has demonstrated superior performance in terms of mean squared error (MSE), making it a compelling alternative for channel estimation in semantic communication systems \cite{xie2020lite}.

IoT devices often face constraints such as limited memory, low computational power, and short battery life, which pose challenges for deploying complex DL models. To mitigate these limitations, much of the computational load can be offloaded to cloud or edge platforms, where DL models are trained on IoT data before deployment \cite{baghban-2024-IoT}. However, the large size of DL models can introduce high latency and increased power consumption, impacting their performance in resource-constrained environments. Pruning and quantization offer efficient solutions for creating lightweight models capable of addressing the resource limitations of IoT devices in SC \cite{xie2020lite}. Pruning reduces model complexity by removing unnecessary parameters, thereby conserving memory and processing capacity \cite{EDROPOUT2022}. On the other hand, quantization decreases the precision of model weights, reducing memory requirements while maintaining acceptable performance levels \cite{shi-2024Q}.

\subsection{Related Work} 
DL-based joint JSCC schemes have gained significant attention for wireless image transmission surpassing traditional image codecs and channel codes in both efficiency and performance \cite{8723589, 9438648, xie2021task, 9066966}. Early works, such as \cite{8723589}, introduced CNN-based JSCC models, demonstrating superior robustness under low signal-to-noise ratio (SNR) conditions. Advancements like attention-based JSCC \cite{9438648} and adaptive bottleneck-guided compression \cite{sun2023adaptive} further improved transmission efficiency, while generative models such as InverseJSCC and GenerativeJSCC \cite{erdemir2023generative} leveraged pre-trained StyleGANs for enhanced perceptual quality.

In multi-user and multi-device settings, collaborative JSCC \cite{lo2023collaborative} utilized channel-aware deep neural networks (DNNs) to enhance image retrieval in additive white Gaussian noise (AWGN) and Rayleigh fading conditions. ResNet-based task-oriented JSCC \cite{xie2021task} was developed for multimodal data, while autoencoder-driven aerial image classification \cite{9796572} and SNR-adaptive JSCC \cite{9414037} demonstrated improved adaptability. Recently, transformer-based models like WITT \cite{10094735} employed the Swin Transformer to optimize global dependencies in HR image transmission.

DL has become a key enabler in communication systems that improves modulation and channel estimation \cite{o2017introduction, 8640815, bai-2020CE, peng-2022CE, gumus-2023, li-2022CE}. In orthogonal frequency division multiplexing (OFDM) systems, DNNs have been used for channel estimation and data demodulation, improving performance under time-varying conditions \cite{gumus-2023}. In \cite{li-2022CE}, a lightweight transformer-based estimator was introduced, which combined CNNs with resource-efficient transformers to improve accuracy while reducing complexity. Furthermore, a CNN deactivating CNN for multiple-input multiple-output (MIMO) uplink systems \cite{zeng2019toward} demonstrated better performance with lower computational cost compared to traditional methods.

Model compression techniques, including pruning and quantization, are increasingly integrated into JSCC and semantic communication frameworks. Works such as DeLighT-based JSCC \cite{jia2023lightweight} and energy-based pruning \cite{EDROPOUT2022} demonstrated significant reductions in model size with minimal accuracy loss. Additionally, CNN pruning methods \cite{THINEt2022} and low-precision vision transformers \cite{shi-2024Q} have further optimized computational efficiency, making transformer-based architectures feasible for resource-constrained IoT devices.
\subsection{Motivations and Contributions} 
Earlier CNN-based JSCC models struggle to capture essential semantics effectively leading to performance degradation as image dimensions increase \cite{8723589, 9438648, 9066966}. Vision transformers, with their advanced attention mechanisms, enable a deeper understanding of high-level semantic content focusing on the meaningful elements critical for robust and efficient communication \cite{khan2022transformers}. Besides, DL-based JSCC techniques have to dynamically adapt to varying SNR conditions, where discrepancies between training and deployment environments can significantly impact performance \cite{8723589, burth2020joint}. While existing methods have introduced channel-dependent modules to mitigate such disparities \cite{10094735,9438648}, these often achieve low gain, especially in rapid channel state fluctuations. Additionally, channel estimation for JSCC image transmission remains insufficiently explored, particularly in developing intelligent solutions adapting to high network dynamics. Unlike conventional approaches that rely on specific scenarios, there is significant potential to create adaptive techniques that enhance resilience and efficiency across diverse channel conditions \cite{10094735, 8723589, 9438648}. Furthermore, lightweight models are essential for ensuring efficient deployment in resource-constrained environments, such as IoT devices, where computational and energy limitations should be carefully considered.

Motivated by the above design issues, our paper introduces a novel JSCC architecture called Swin semantic image transmission (SwinSIT) for wireless image transmission. With a focus on HR images, the SwinSIT adopts Swin transformers for broad global semantic information extraction and reconstruction of image patches on both the transmitter and receiver sides. In addition, an SNR-aware module is proposed to enhance the semantics captured by the Swin transformer on both the transmitter and receiver sides. By adapting to dynamic network conditions, the SNR-aware module ensures robust image transmission maintaining high-quality semantics despite signal fluctuations. Also, SwinSIT contains a CNN-aided channel estimator and compensator (CEAC) to accurately estimate fading coefficients and remove their effect by exploiting the image denoising CNN (DnCNN) architecture. Furthermore, a joint pruning and quantization process is presented to compress the proposed SwinSIT model sustaining its favorable performance to fulfill the resource constraints of IoT devices. 

To evaluate the effectiveness of our proposal, we conducted a comprehensive comparison against several recent benchmarks for image transmission, considering various design perspectives. These benchmarks include the transformer-based WITT model, the CNN-based Deep JSCC, and the conventional BPG+LDPC scheme, as presented in \cite{10094735}, \cite{8723589}, and \cite{BPG, 8316763}, respectively. The evaluation metrics include PSNR, multi-scale structural similarity index (MS-SSIM), and a human perceptual metric. Simulations show that the proposed SwinSIT outperforms the conventional schemes across these metrics even with compression.

The main contributions of our paper are summarized as follows
\begin{itemize}
\item Exploiting hierarchical semantic maps constructed by the vision transformers, an encoder/decoder architecture is designed for semantic image transmission relying on the Swin transformer for the Rayleigh fading channel.
\item A novel CNN-based SNR-aware module is introduced to enhance the semantic map extracted by the encoder, as well as its noisy counterpart received at the decoder, ensuring robust performance in dynamic SNR conditions. 
\item A new CNN-based CEAC utilizing the DnCNN network is presented to estimate fading coefficients and mitigate the effects of channel fading. 
\item A joint pruning and quantization process is introduced to compress the SwinSIT model for efficient IoT resource utilization while maintaining high performance.

\item The PSNR and MS-SSIM performance of SwinSIT is evaluated through extensive simulations, and compared with traditional benchmark models to confirm its adaptability and durability by covering both low and high-quality images.
\end{itemize}

The rest of this paper is organized as follows: In Section \ref{Sec_SM_PF}, the system model is presented in addition to the problem formulation. Section \ref{Sec_Arch} provides an overview of the proposed SwinSIT architecture with a detailed description of its blocks including the proposed compression process. In Section \ref{Sec_Sim}, the performance of the proposed model is evaluated and compared to conventional benchmarks. Finally, this paper is concluded in Section \ref{Sec_Conc}.  
\section{System model and problem formulation} \label{Sec_SM_PF}
\subsection{System Model} \label{SubSec_SM}
\begin{figure}[th]
\centering
\centerline{\includegraphics[width=\columnwidth]{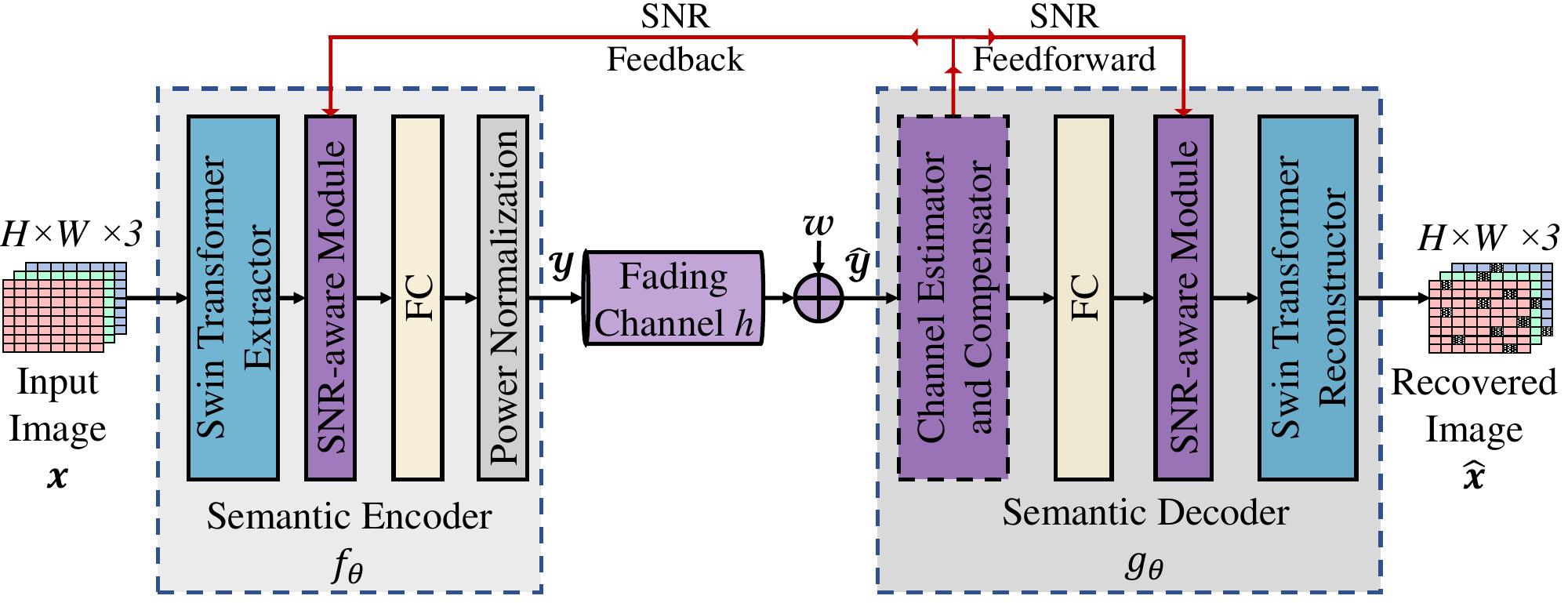}}
\caption{The model of the proposed wireless image transmission system.}
\label{fig:SM}
\end{figure}

In Fig. \ref{fig:SM}, a point-to-point wireless SC system for image transmission is presented based on the Swin transformer encoder-decoder design. The system model includes two Swin transformer blocks, one on the transmitter side as a semantic extractor and the other on the receiver side as an image reconstructor. Besides, assuming exact knowledge of SNR at the receiver, two SNR-aware modules are attached. The first one produces an enhanced version of the semantic map extracted by the Swin transformer at the encoder, while the second produces an improved version of the noisy semantic map received at the decoder. Also, a CEAC determines the channel parameter $h$ for data demodulation at the receiver. Two fully connected (FC) layers are incorporated, one at the encoder and the other at the decoder, to transform and refine the extracted semantics. Additionally, a power normalization layer is added at the end of the encoder to optimize the transmission. The detailed description for all parts and layers is explained in Section \ref{Sec_Arch}.

We assume an input image $\bm{\mathit{x}} \in \bm{\mathbb{R}}^n$ of size $n = H\times W \times 3$, where $H$ and $W$ represent the height and the width, respectively and $3$ indicates the number of RGB channels. The encoder encodes the real-valued image vector $\bm{\mathit{x}}$ of size $n$ with the aid of the SNR feedback to produce a complex-valued vector $\bm{\mathit{y}}$ of size $k$ to be sent through the Rayleigh channel. The encoding function $f_\theta:\bm{\mathbb{R}}^n \times \bm{\mathbb{R}}\rightarrow{\bm{\mathbb{C}}^k}$ can be written as
\begin{equation}
\bm{\mathit{y}} = f_{\bm{\theta}}(\bm{\mathit{x}}, \text{SNR} ) 
\label{eq:SM_Tx}
\end{equation}
where $\bm{\theta}$ is the set of parameters for the semantic encoder. The model considers the SNR value fed back from the receiver to improve the training process performance.

The vector $\bm{\mathit{y}}$ is transmitted through a fading channel and the channel output vector $\hat{\bm{\mathit{y}}} \in {\bm{\mathbb{C}}^k}$ can be denoted as

\begin{equation}
\hat{\bm{\mathit{y}}} = h \bm{\mathit{y}} + \bm{\mathit{w}},
\label{eq:SM_Ray}
\end{equation}
where $h$ represents the Rayleigh slow fading channel coefficient, and $\bm{\mathit{w}}$ indicates the complex Gaussian noise vector with zero mean and covariance $\sigma^2 \bm{\mathit{I}}$. Also, the receiver utilizes the decoding function $g_\phi:\bm{\mathbb{C}}^k \times \bm{\mathbb{R}}\rightarrow{\bm{\mathbb{R}}^n}$ as
\begin{equation}
\hat{\bm{\mathit{x}}} = g_{\bm{\phi}}(\hat{\bm{\mathit{y}}}, \text{SNR}),
\label{eq:SM_Rx}
\end{equation}
where the parameter set $\bm{\phi}$ corresponds to the set of semantic decoder's parameters. 

\subsection{Problem formulation} \label{SubSec_PF}
Following \cite{8723589, 9066966, 9438648}, we consider the image size and the channel input size as $n$ and $k$, respectively. Accordingly, we define code rate $R=\frac{k}{n}$. Given a specific $R$, our main objective is to train the proposed model in an end-to-end fashion to identify the optimal semantic encoder and decoder parameters $\bm{\theta}^*$ and $\bm{\phi}^*$ such that the expected distortion is minimized. In this context, our problem can be formulated as
\begin{equation}
(\bm{\theta}^*, \bm{\phi}^*)=\arg\min_{\substack{\bm{\theta},\bm{\phi}}} {\bm{\mathbb{E}}}_{p(\text{SNR})}{\bm{\mathbb{E}}}_{p(\bm{\mathit{x}},\hat{\bm{\mathit{x}}})}d(\bm{\mathit{x}}, \hat{\bm{\mathit{x}}})  ,
\label{eq:SM_problem}
\end{equation}
where $p(\text{SNR})$ characterizes the probability distribution of the SNR, $p(\bm{\mathit{x}},\hat{\bm{\mathit{x}}})$ describes the joint probability distribution of the transmitted image ${\bm{\mathit{x}}}$ and the recovered image $\hat{{\bm{\mathit{x}}}}$, with ${\bm{\mathbb{E}}}_{p(SNR)}$ and ${\bm{\mathbb{E}}}_{p(\bm{\mathit{x}},\hat{\bm{\mathit{x}}})}$ representing the corresponding expectations. The distortion between $\bm{\mathit{x}}$ and $\hat{\bm{\mathit{x}}}$ is measured as follows
\begin{equation}
d(\bm{\mathit{x}}, \hat{\bm{\mathit{x}}})=\frac{1}{n}\sum_{i=1}^n (x_i-\hat{x}_i)^2,
\label{eq:SM_distorsion}
\end{equation}
where $x_i$ being the $i^{th}$ element of $\bm{\mathit{x}}$.
\section{SwinSIT Architecture} \label{Sec_Arch}
\begin{figure*}[!ht]
\centering
\begin{center}
\includegraphics[width=1.5\columnwidth]{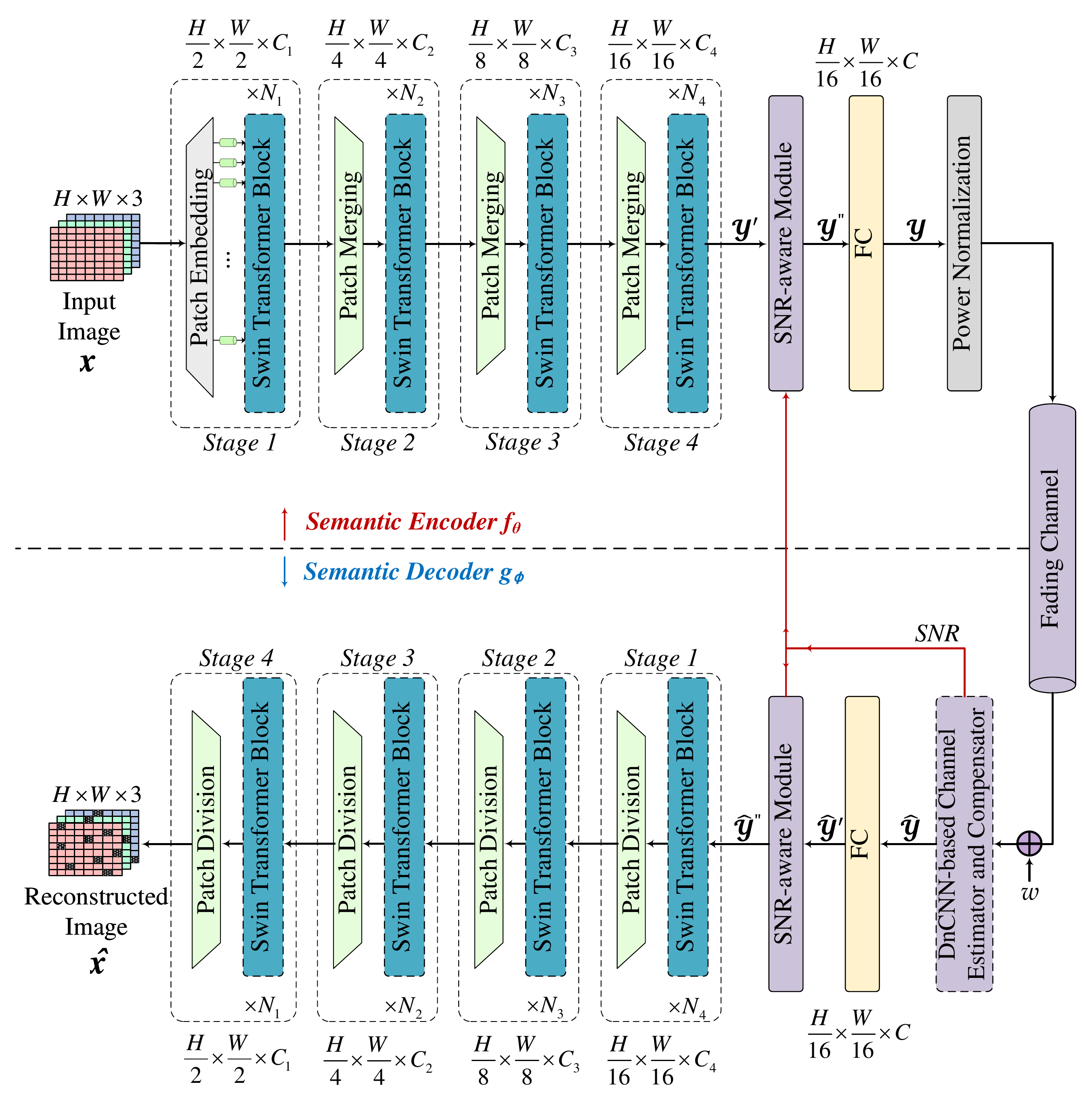} 
\caption{The proposed SwinSIT architecture.}
\label{Fig2_Arch}
\end{center}
\end{figure*}
The description of the overall SwinSIT architecture consisting of an encoder and a decoder based on the Swin transformer is depicted in Fig. \ref{Fig2_Arch}. The main task of the encoder is semantic extraction and transmission, while that of the decoder is symbol reception, channel estimation and compensation, and image reconstruction. The encoder comprises a series of stages, where each stage contains either patch embedding or patch merging as well as a Swin transformer block. Stage $1$ at the encoder includes patch embedding that divides the input RGB image into  $l_1 = \frac{H}{2} \times \frac{W}{2}$ non-overlapping patches with a chosen number of channels $C_1$. The output patches are fed to the Swin transformer block to apply the shifted window multi-head self-attention mechanism as detailed in the state-of-the-art work in\cite{liu2021Swin}. The remaining stages include patch merging modules as downsamplers. As depicted by the SwinSIT encoder design, the number of patches is reduced to $l_2 = \frac{H}{4} \times \frac{W}{4} \text{, } l_3 = \frac{H}{8} \times \frac{W}{8}\text{, and } l_4 = \frac{H}{16} \times \frac{W}{16} $ at stage 2, stage 3, and stage 4, respectively with chosen channels $C_2, C_3, \text{and }C_4$. During these stages, the patch merging outputs are fed to the Swin transformer blocks for a better semantic map. It is worth mentioning that the number of stages depends on the transmitted image resolution, which will be shown in the simulation settings. Also, an SNR-aware module comes after the four stages to perform a double-stage enhancement for the extracted semantic map. Besides, an FC layer is added to reshape the output of channels to a desired number $C$ for achieving a certain $R$ during the training. Furthermore, power normalization is applied to satisfy the power constraint before transmission.

On the other hand, the first block in the decoder is the CEAC, which estimates the fading channel $h$ and compensates for the fading effect.
Then, an FC layer is added to reshape the received data with $C_4$ channels followed by another SNR-aware module for noisy semantic map enhancements. Next, four stages are inserted, where each stage consists of a Swin transformer block and a patch division module. Each Swin transformer block works as an image reconstructor, whereas each patch division module serves as an upsampler for reversing the downsampling process at the encoding. Finally, $l_1 = \frac{H}{2} \times \frac{W}{2}$ patches are produced before the last patch division module in stage $4$ to reconstruct the received image $\hat{\bm{\mathit{x}}}$.
\subsection{SNR-aware Module} 
Here, we give a detailed description of the proposed SNR-aware module inspired by the squeezing-and-excitation (SE) network proposed in \cite{hu2018squeeze} to further exceed the channel modnet performance limits proposed in \cite{10094735}. This module is utilized to scale the semantic map according to their contributions to the image reconstruction process relying on the SNR determined at the receiver. The SNR-aware module at the encoder side handles the semantics extracted from different stages of the transformer blocks. Also, the SNR-aware module at the decoder enhances the semantics of the received signal. This allows them to adjust their output based on the prevailing channel conditions to ensure adaptability. 
\begin{figure*}[t]
\centering
\centerline{\includegraphics[width=1.6\columnwidth]{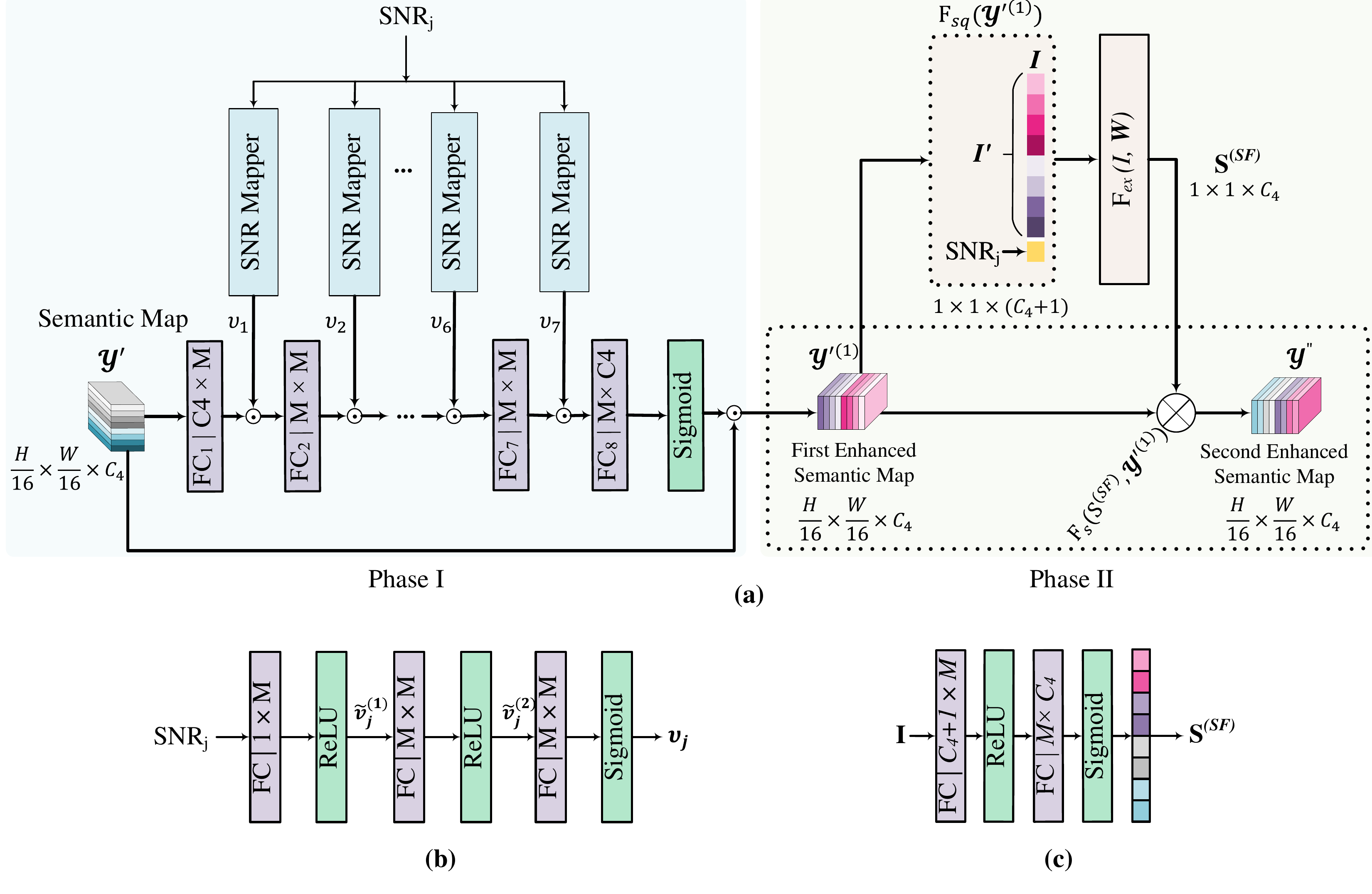}}
\caption{(a) The proposed SNR-aware module (b) SNR mapper block (c) excitation block.}
\label{fig_SNR}
\end{figure*}
The design of an SNR-aware module is shown in Fig. \ref{fig_SNR}(a) which consists of two phases. In Phase I, the semantic map ${\bm{\mathit{y}}}^\prime$ is scaled at the encoder side based on $8$ FC layers and $7$ SNR mappers, to produce the first version of the enhanced semantic map. Here, a three-layered FC network with the current channel SNR input SNR$_j$ makes up the SNR mapper block as detailed in Fig. \ref{fig_SNR}(b). This SNR mapper block creates an $M$-dimensional vector $\bm{\mathit{v}}_{j}$ $\big(j\in{\{1, 2, ..., 7\}\big)}$ based on the receiver SNR feedback as
\begin{IEEEeqnarray}{lrC}
\IEEEyesnumber\label{eq:SM_module} \IEEEyessubnumber*
\tilde{\bm{\mathit{v}}}_j^{(1)}=\text{ReLU}(\bm{W}^{(1)} \cdot \text{ SNR}_j+\bm{b}^{(1)}), \label{}\\
\tilde{\bm{\mathit{v}}}_j^{(2)}=\text{ReLU}(\bm{W}^{(2)} \muldot \textbf{}\tilde{\bm{\mathit{v}}}_j^{(1)}+\bm{b}^{(2)}), \label{}\\
\bm{\mathit{v}}_j=\text{Sigmoid}(\bm{W}^{(3)} \muldot \textbf{}\tilde{\bm{\mathit{v}}}_j^{(2)}+\bm{b}^{(3)}), \label{}
\end{IEEEeqnarray}
where the rectified linear unit (ReLU) and sigmoid are the activation functions, $\bm{W}^{(k)}$ and $\bm{b}^{(k)}$ represent the parameters of the linear function of the $k^{th}$ FC layer and their biases, respectively, and $\muldot$ denotes the element-wise multiplication.

After producing $\bm{\mathit{v}}_j$ by the SNR mapper block, an element-wise multiplication operation is performed to the first 7 FC layers output of Phase I and $\bm{\mathit{v}}_j$. Then, the result is forwarded to the subsequent FC layer in the same phase. The 8$^{th}$ FC layer's decision is then disseminated into an area with the same spatial extent as ${\bm{\mathit{y}}}^\prime$. Different $\bm{\mathit{v}}_j$ pay attention to the SNR value in our SNR-aware proposal so that the first enhanced semantic map is produced in a channel-adaptive manner. In Phase II, another enhancement for the semantic map is performed based on an SNR-adaptive version of the SE network. Here, the first enhanced semantic map is received from Phase I to yield the second enhanced semantic map as shown in Fig. \ref{fig_SNR}(a). 

Let ${\bm{\mathit{y}}}^{\prime(1)} = [{\bm{\mathit{y}}}_{1}^{\prime(1)}, {\bm{\mathit{y}}}_2^{\prime(1)}, ..., {\bm{\mathit{y}}}_{C_4}^{\prime(1)}] \in \bm{\mathbb{R}}^{H^\prime \times W^\prime \times C_4}$ stand for the first enhanced semantic map, where $C_4$ represents the number of semantics and $H^\prime \times W^\prime$ indicates the size of each semantic. Also, let ${\bm{\mathit{y}}}^{\mydprime} = [{\bm{\mathit{y}}}_1^{\mydprime}, {\bm{\mathit{y}}}_2^{\mydprime}, ..., {\bm{\mathit{y}}}_{C_4}^{\mydprime}] \in \bm{\mathbb{R}}^{H^\prime \times W^\prime \times C_4}$ denote the scaled semantics obtained by the two phases. Accordingly, the SNR-adaptive SE of Phase II employs a global average pooling squeezing function F$_{sq}(.)$ to analyze ${\bm{\mathit{y}}}^{\prime(1)}$. Here, the SNR$_j$ and semantic information $\bm{I}^\prime$ are instances of the squeezing function F$_{sq}(.)$ producing $\bm{I}$. In most cases, convolutional filters restricted to a local receptive area are able to extract image semantics. Therefore, especially when performing semantic extraction with a tiny kernel size, these semantics typically cannot sense the information outside this region. However, by averaging the elements $u_{jk}^{(i)}$ of ${\bm{\mathit{y}}}_i^{\prime(1)}$, F$_{sq}(.)$ can extract global information as
\begin{equation}
{\mathit{I}}_i^\prime=F_{sq}({\bm{\mathit{y}}}_i^{\prime(1)})=\frac{1}{{H^\prime \times W^\prime}}\sum_{\zeta=1}^{H^\prime}\sum_{\psi=1}^{W^\prime}{u_{jk}}.
\label{eq:GAP}
\end{equation}

Then, the context information $\bm{I}$ is generated by concatenating the output of $F_{\text{sq}}(.)$, denoted as $\mathit{I}_i^\prime$, with the SNR, resulting in
\[
\bm{I} = (\text{SNR}_j, \mathit{I}_1^\prime, \mathit{I}_2^\prime, \dots, \mathit{I}_{C_4}^\prime).
\]
Afterward, a FC NN receives the context information and generates a scaling factor through the excitation process $F_{\text{ex}}(\cdot, \cdot)$. Based on the context information $\bm{\mathit{I}}$, we apply an excitation operation to estimate the scaling factor $\bm{S}^{(\text{SF})}$, where $F_{\text{ex}}(\bm{I}, \bm{W})$ represents an NN composed of two FC layers, designed to avoid excessively increasing complexity. The first FC layer applies a ReLU activation function, while the second FC layer uses a sigmoid activation function. Accordingly, the scaling factor $\bm{S}^{(\text{SF})}$ can be expressed as
\begin{equation}
\bm{S}^{(\text{SF})} = F_{\text{ex}}(\bm{I}, \bm{W}),
\end{equation}
where $\bm{W}$ denotes the set of trainable parameters within the excitation network.

\begin{align}
\bm{\textbf{S}^{(SF)}}&=\text{F}_{ex}(\bm{I},\bm{W})\nonumber\\
&=\text{Sigmoid}\big({\bm{W}_2^{\bm{(SF)}}} \text{ReLU} ( {\bm{W}_1^{\bm{(SF)}}}\bm{I}+\bm{b}_1^{\bm{(SF)}} )\nonumber\\
&+\bm{b}_2^{\bm{(SF)}} \big),
\label{eq:S_FP}
\end{align}
where $\bm{W}_k^{\bm{(SF)}}$ and $\bm{b}_k^{\bm{(SF)}}$ represent the $k^{th}$ FC layer's weights and biases, respectively ($k$=1, 2). Then. by multiplying the semantics first enhanced semantic map and the scaling factor $\textbf{F}_s$, the second enhanced semantic map ${\bm{\mathit{y}}}^{\mydprime}$ is produced as follows
\begin{align}
{\bm{\mathit{y}}}^{\mydprime}= \textbf{F}_s\big(\bm{\textbf{S}^{(SF)}},{\bm{\mathit{y}}}^{\prime(1)} \big)=\bm{\textbf{S}^{(SF)}} \muldot {\bm{\mathit{y}}}^{\prime(1)}.
\label{eq:FR}
\end{align}
Depending on the SNR conditions, several enhanced semantics are created where the value of each one reflects its importance to image reconstruction at the decoder side. The procedures of our proposed SNR-aware module at the encoder side are detailed in Algorithm \ref{Alg_SNR}. Similarly, on the decoder side, a double enhancement for the channel output $\hat{{\bm{\mathit{y}}}}^\prime$ takes place to produce the second enhanced semantic map $\hat{{\bm{\mathit{y}}}}^{\mydprime}$ relying on the SNR value. 
\begin{algorithm}[tb]
\footnotesize
	\caption{SNR-aware Module}\label{Alg_SNR}
	\begin{algorithmic}[1]
		{
\renewcommand{\algorithmicrequire}{\textbf{Input:}}
                \STATE \textbf{1) Phase I}
                \STATE Reshape the number of channels of the input semantic map from $C_4$ to $M$ using the first FC layer.
                \FOR{$i=0:n$ }
                \STATE Use SNR feedback to create $\bm{\mathit{v}}_j$ using \eqref{eq:SM_module}.
                \STATE Perform element-wise multiplication to fuse the semantics’ map with $\bm{\mathit{v}}_j$.
                \ENDFOR
                \STATE Obtain the first enhanced semantic map.
                \STATE \textbf{2) Phase II}
                \STATE Extract ${\mathit{I}}_i^\prime$ as the global information of ${\bm{\mathit{y}}}^{\prime(1)}$ using \eqref{eq:GAP}.
                \STATE Concatenate the SNR with ${\mathit{I}}_i^\prime$ to form the context information $\bm{I}= (\text{SNR}_j, \mathit{I}_1^\prime, \mathit{I}_2^\prime, ..., {\mathit{I}}_{C_4}^\prime)$.
                \STATE Determine the scaling factor $\bm{\textbf{S}^{(SF)}}$ using \eqref{eq:S_FP}.
                \STATE Transform ${\bm{\mathit{y}}}^{\prime(1)}$ and $\bm{\textbf{S}^{(SF)}}$ to their channel-wise form as ${\bm{\mathit{y}}}_i^{\prime(1)}$ and $S_i^{\text{(SF)}}$, respectively, for $\textbf{ }i=1, 2, ..., C_4$.
			\FOR{$i=0:C_4$ }
			\STATE Perform element-wise multiplication on ${\bm{\mathit{y}}}_i^{\prime(1)}$ and $S_i^{\text{(SF)}}$ using \eqref{eq:FR}.
			\ENDFOR
                \STATE Use the multiplication results to reconstruct the second enhanced semantic map ${\bm{\mathit{y}}}_i^{\mydprime}$.
			\renewcommand{\algorithmicensure}{\textbf{Output:}}
		}
	\end{algorithmic}
\end{algorithm}

\subsection{DnCNN-based channel estimator and compensator} \label{SubSec_Arch_Swin}
\begin{figure}[th]
\centering
\centerline{\includegraphics[width=0.8\columnwidth]{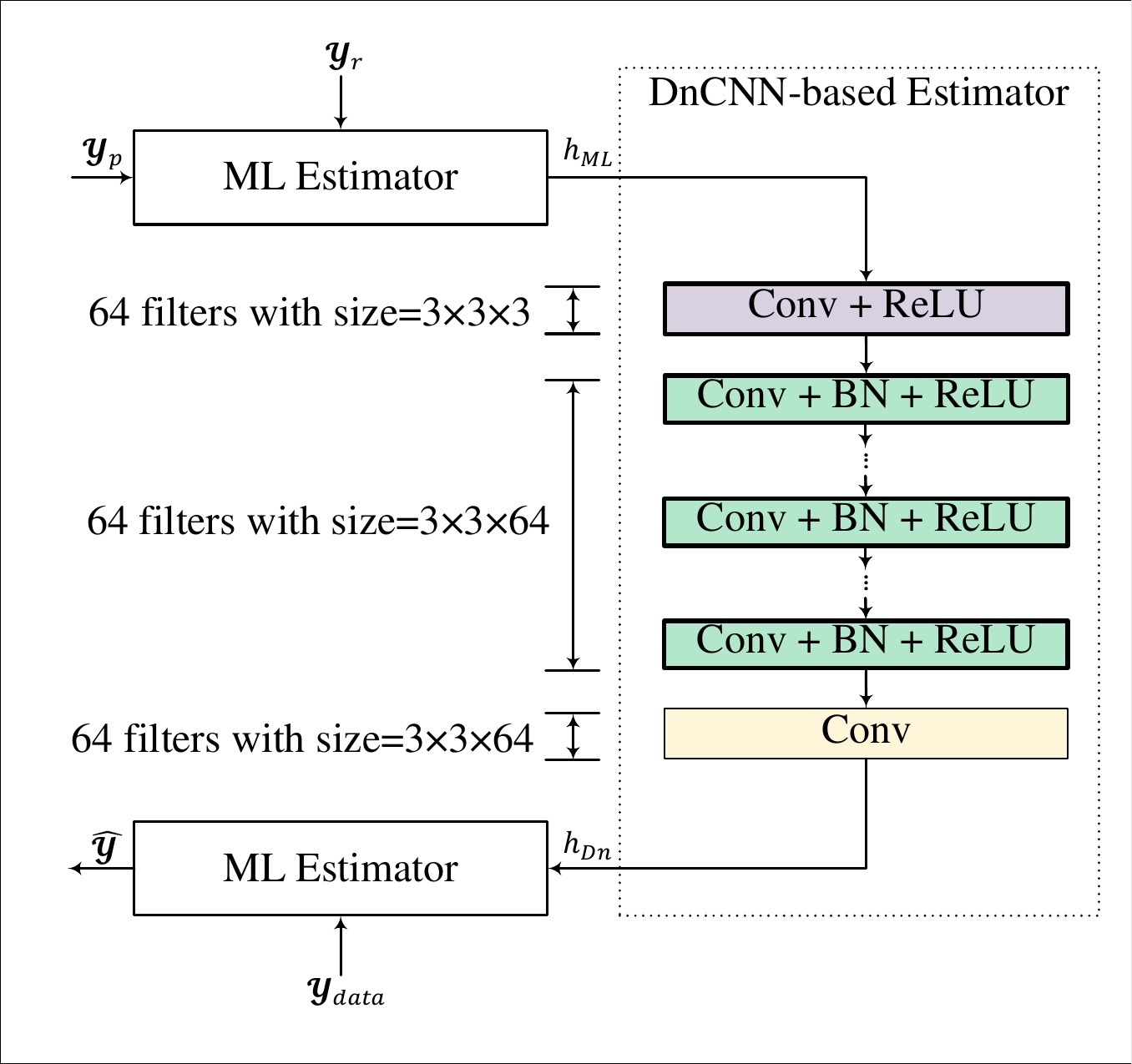}}
\caption{The proposed DnCNN-based channel estimator and compensator.}
\label{fig_DnCNN}
\end{figure}

The DnCNN network \cite{DnCNN} achieved good performance in image denoising tasks by removing the effect of AWGN added to the images. Thus, we utilize the DnCNN network to enhance the channel estimate. Our DnCNN-based CEAC is divided into two steps. First, the pilot signals ${\bm{\mathit{y}}}_p$ are transmitted through the fading channel to apply the ML estimator for an initial estimate of the channel parameter as \cite{4518455}
\begin{equation}
h_{ML} = \frac{ {\bm{\mathit{y}}}_p^H {\bm{\mathit{y}}}_r }{{\bm{\mathit{y}}}_p^H {\bm{\mathit{y}}}_p},
\label{eq:ML}
\end{equation}
where $(.)^H$ defines the complex conjugate transpose, and ${\bm{\mathit{y}}}_r$ represents the received signal. Generally speaking, the ML estimate of the fading channel consists of the true value and a quantity that depends on the AWGN. Therefore, we can rewrite the ML estimate as $
h_{ML} = h + \frac{ {\bm{\mathit{y}}}_p^H {\bm{\mathit{w}}} }{{\bm{\mathit{y}}}_p^H {\bm{\mathit{y}}}_p}$.
 The detailed architecture of DnCNN comprises three layers as depicted in Fig. \ref{fig_DnCNN}. The refinement process of the initial ML channel parameter can be expressed as $\textbf{H}_{Dn} = \text{DnCNN}(\textbf{H}_{ML})$.
Then, the DnCNN network is trained based on the loss function ${\mathcal{L}}(\textbf{H}_{Dn}, \textbf{H}) = \frac{1}{2}||\textbf{H}_{Dn}-\textbf{H}||_F^2$.

\subsection{SwinSIT Compression}
This subsection details the joint pruning and quantization approach for compressing the SwinSIT model, optimizing it for IoT devices by reducing its size and complexity while preserving performance.
\begin{figure}[tb]
\centering
\centerline{\includegraphics[width=0.8\columnwidth]{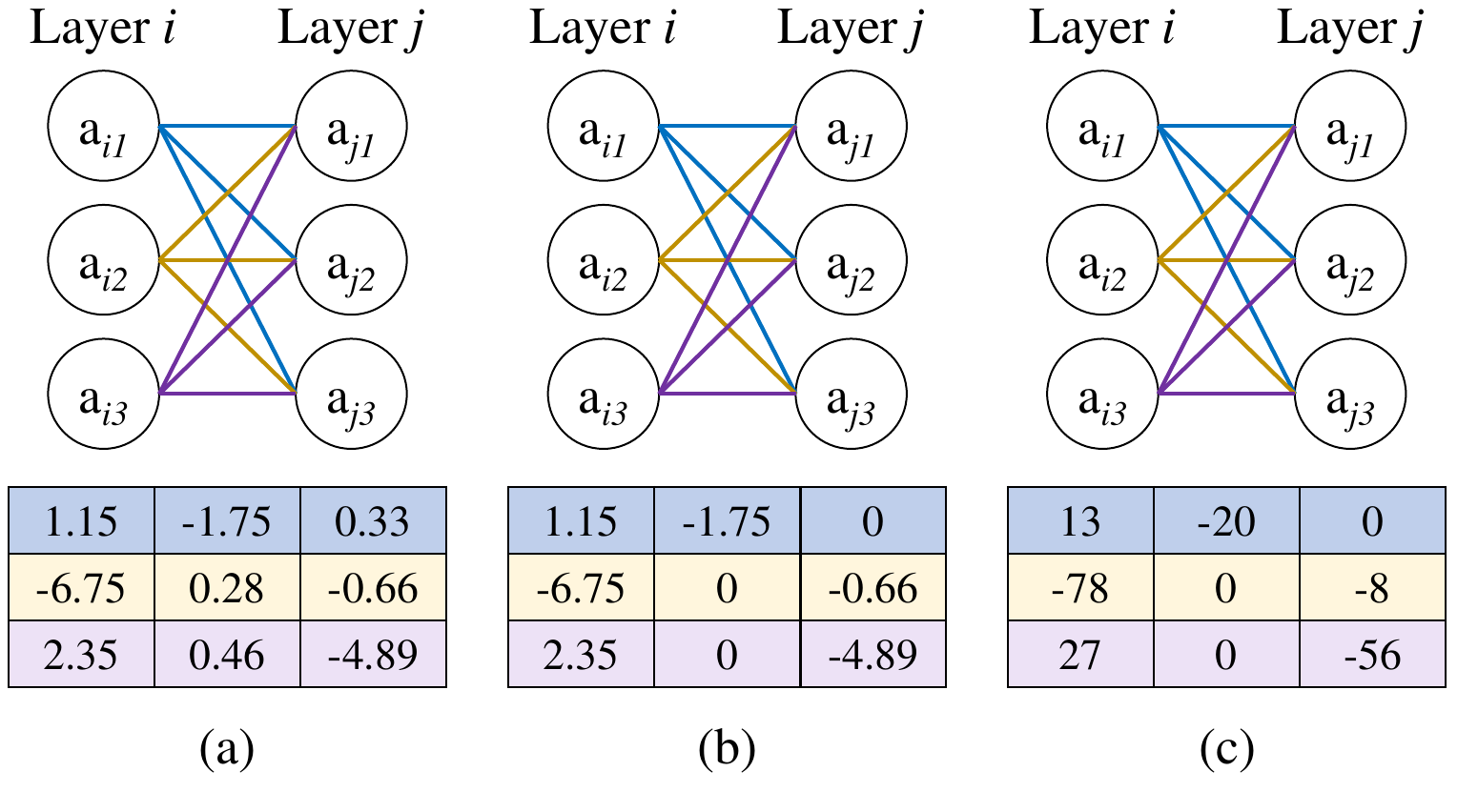}}
\caption{The proposed joint pruning-quantization scheme (a) original weights (b) weights after pruning (c) quantized weights. }
\label{fig_PandQ}
\end{figure}
Our SwinSIT model mainly comprises dense layers. Thus, to illustrate the proposed joint pruning-quantization technique, Fig. \ref{fig_PandQ} highlights an example of the weight transformation between two selected dense layers $i$ and $j$. Initially, the weights connecting layer $i$ to layer $j$ (Fig. \ref{fig_PandQ}(a)) are represented as a dense matrix where every node in layer $i$ links to all corresponding nodes in layer $j$. The pruning process (Fig. \ref{fig_PandQ}(b)) eliminates ineffective weights based on a predefined threshold, forming a sparse NN. Subsequently, quantization is applied (Fig. \ref{fig_PandQ}(c)), reducing the precision of the pruned weights, for example, from FP$32$ to INT$8$ achieving a compact and efficient model. Specifically a detailed description of the proposed joint pruning-quantization scheme is provided below.
\subsubsection{SwinSIT pruning} The pruning strategy focuses on dense layers, which form the majority of the SwinSIT architecture, by targeting and removing insignificant weights. For a pre-trained SwinSIT model with $N$ layers, the weight connecting neuron $l$ in layer $j$ to neuron $m$ in layer $i$ is denoted as $W^{(i)}_{l,m}$. A pruning threshold $\gamma$ is introduced and weights are updated as
\begin{equation}
W^{(i)}_{l,m} = 
\begin{cases}
W^{(i)}_{l,m},& |W^{(i)}_{l,m}| > \gamma,\\
0,              & \text{otherwise}
\end{cases}.
\label{eq:Prun}
\end{equation}
The threshold $\gamma$ is determined based on a sparsity ratio \cite{xie2020lite}. After pruning, the model undergoes fine-tuning to recover any performance loss and ensure robust accuracy.
\subsubsection{SwinSIT quantization} The quantization process targets both weights and activations, transforming the SwinSIT weights to low-precision integers (e.g., INT$8$). The quantized weights are obtained as
\begin{equation}
\Tilde{\bm{W}}^{(i)}_{l,m} = \text{round}( {\alpha_{w} ({\bm{W}}^{(i)}_{l,m}-\text{min}({\bm{W}}^{(i)})) } ),
\label{eq:W_QFn}
\end{equation}
where the scaling factor $\alpha_{w}$ maps the dynamic range of FP$32$ to the $m_q$-bit integer representation as
\begin{equation}
\alpha_{w}=\frac{2^{m_q}-1}{\text{max}({\bm{W}}^{(i)})-\text{min}({\bm{W}}^{(i)})}.
\label{eq:SF_ALPHA_w}
\end{equation}

When performing matrix multiplications, it is crucial to account for potential accumulator overflow caused by the limited dynamic range of integer formats. For instance, an accumulator might require a higher bit width (e.g., INT$32$) when combining INT$8$ weights and activations. Additionally, dynamic changes in input data can result in activation outliers, which significantly affect dynamic range. To address this, an exponential moving average is used to smooth the quantization range of activations given as
\begin{equation}
a^{(i)}_{min}(t+1)=(1-\beta)a^{(i)}_{min}(t)+\beta \text{ min}(\bm{A}^{(i)}(t)),
\label{eq:x_min}
\end{equation}
\begin{equation}
a^{(i)}_{max}(t+1)=(1-\beta)a^{(i)}_{max}(t)+\beta \text{ max}(\bm{A}^{(i)}(t)).
\label{eq:x_max}
\end{equation}
Here, $\bm{A}^{(i)}(t)$ represents the activation output at layer $i$ for batch $t$, and $\beta \in [0,1)$ controls the influence of new data on the range.

To ensure stability, activation quantization outputs are clamped as follows
\begin{equation}
\Tilde{\bm{A}}^{(i)} = \text{clamp} ( \text{round}( {\alpha_{a} ({\bm{A}}^{(i)}-a^{(i)}_{min} ) } ); -M,M ),
\label{eq:Act_QF}
\end{equation}
where $\alpha_{a}$ is the scaling factor defined as
\begin{equation}
\alpha_{a}=\frac{2^{m_q}-1}{a^{(i)}_{max}-a^{(i)}_{min}},
\label{eq:SF_ALPHA_a}
\end{equation}
\begin{equation}
\text{clamp}(\bm{A}^{(i)};-T,T)=\text{min}( {\text{max}(\bm{A}^{(i)},-T),T }),
\label{eq:clamp}
\end{equation}
where the clamp function limits activation outputs to the range $[-T, T]$ with $T = 2^{m_q} - 1$. This ensures precise quantization while mitigating the effects of outliers.

As depicted in Algorithm \ref{Alg_P&Q}, the process of SwinSIT joint pruning and quantization comprises three distinct steps: 1) weights pruning, 2) weights quantization, and 3) activations quantization. In step $1$, the pruning of the unnecessary weights among layers can be directly performed following \eqref{eq:Prun}. After that, the remaining weights of each layer can be quantized using \eqref{eq:W_QFn} in step $2$. In step $3$, a calibration procedure is implemented by executing several calibration batches to acquire activation statistics. In each batch, the values $a^{(i)}_{min}(t+1)$ and $a^{(i)}_{max}(t+1)$ are updated based on the activation statistics gathered. It's worth noting that these quantization procedures may result in marginal accuracy reductions. To mitigate this, quantization-aware training becomes necessary to retrain the model, minimizing the loss of accuracy \cite{9398534}. Given that the rounding operation lacks differentiability, a straight-through estimator is employed to estimate the gradient of quantized weights during back-propagation \cite{bengio2013estimating}.
\begin{algorithm}[tb]
    \caption{Joint Pruning and Quantization Algorithm}\label{Alg_P&Q}
    \begin{algorithmic}[1]
        {
        \footnotesize
        \renewcommand{\algorithmicrequire}{\textbf{Input:}}
        \REQUIRE SwinSIT weights $\bm{W}$, the pruning threshold $\gamma$, the desired quantization low-precision $m_q$, the calibration dataset $\mathcal{D}$.
        \STATE \textbf{Step 1: SwinSIT weights pruning}
        \STATE Determine the overall number of connections $N_w$.
        \STATE Sort these connections ascending.
        \FOR{$n=1:1:N_w$}
            \STATE Perform weights pruning using \eqref{eq:Prun}.
        \ENDFOR
        \STATE Perform fine-tuning for the pruned model.
        \STATE \textbf{Step 2: SwinSIT weights quantization}
        \STATE Determine the number of remaining connections after pruning $N_{p}$.
        \FOR{$n=1:1:N_p$}
            \STATE Determine weights' range in terms of $\text{min}\big({\bm{W}}^{(i)}\big)$ and $\text{max}\big({\bm{W}}^{(i)}\big)$.
            \STATE Perform weights quantization following \eqref{eq:W_QFn} producing quantized weights $\Tilde{\bm{W}}^{(i)}$.
        \ENDFOR
        \STATE \textbf{Step 3: SwinSIT activations quantization}
        \STATE Determine the overall number of activations $N_a$.
        \FOR{$t=1:1:\mathcal{D}$}
            \FOR{$n=1:1:N_a$}
                \STATE Refresh the activation dynamic range by updating $a^{(i)}_{min}(t+1)$ and $a^{(i)}_{max}(t+1)$ using \eqref{eq:x_min} and \eqref{eq:x_max}, respectively.
            \ENDFOR
        \ENDFOR
        \STATE Perform activations quantization following \eqref{eq:Act_QF} producing quantized activations $\Tilde{\bm{A}}^{(i)}$.
        \STATE Utilize the STE to perform fine-tuning for the compressed model.
        \renewcommand{\algorithmicensure}{\textbf{Output:}}
        \ENSURE Return SwinSIT pruned and quantized weights $\Tilde{\bm{W}}^{(i)}$, and quantized activations $\Tilde{\bm{A}}^{(i)}$.
        }
    \end{algorithmic}
\end{algorithm}
\section{Simulation Results} \label{Sec_Sim}
\subsection{Simulation Setup} \label{SubSec_SimSet}
The performance of the proposed SwinSIT model is investigated against the conventional source-channel coding which adopts the BPG codec \cite{BPG} and LDPC \cite{8316763} with a block length of $6144$ bits. Besides, we compare with the Deep JSCC model in \cite{8723589} that design both the encoder and decoder as two CNNs. Moreover, to ensure a fair comparison, we employ the WITT scheme \cite{10094735} based on the Swin transformer which considers the SNR. The WITT model incorporates the channel modnet module for one-phase semantic map enhancement in contrast to the proposed SwinSIT that performs two-phase enhancements. Notably, we assume the perfect channel state information for the BPG$+$LDPC scheme, whereas the least squares method is assumed for the channel estimation of WITT and Deep JSCC models.

The simulations evaluate multiple variants of the SwinSIT model, which include SwinSIT without channel estimation, and SwinSIT SNR-unaware. SwinSIT without channel estimation retains the SNR-aware module but excludes the DnCNN-based CEAC, and SwinSIT SNR-unaware represents the model without both the SNR-aware module and the DnCNN-based CEAC.

In alignment with \cite{10094735,yang-2022,9953099}, the SNR range is set to be $[1$dB, $13$dB] throughout the experiments. We adopt the PSNR and MS-SSIM to assess model performance. To enhance readability, MS-SSIM values are presented in dB as $\text{MS-SSIM (dB)}=-10\text{log}(1-\text{MS-SSIM})$. We aim to investigate our SwinSIT model for the transmission of low-resolution (LR) and HR images as in \cite{10094735}. Their details are shown in Table \ref{tab_train}. Regarding the LR setup, we incorporate the CIFAR10 dataset \cite{krizhevsky2009learning} for training and testing comprising $50000$ and $10000$ images, respectively. For the HR setup, we utilize the DIV2K dataset \cite{agustsson2017ntire} and the Kodak dataset \cite{Kodak} for training and testing, respectively. For the training process, images undergo random cropping to create patches with size $256 \times 256$. For the CIFAR10 dataset, two Swin transformer stages are considered as $[N_1, N_2]=[2, 4]$ with $[C_1, C_2]=[128, 256]$ with the batch size of $128$. On the other hand, for HR images, four Swin transformer stages are employed as $[N_1, N_2, N_3, N_4]=[2, 2, 6, 2]$ with $[C_1, C_2, C_3, C_4]=[128, 192, 256, 320]$ with the batch size of $16$. Notably, we follow \cite{10094735} for the aforementioned hyper-parameters. Pytorch is utilized to implement the simulations of this paper relying on a single NVIDIA GeForce RTX 4070 TI GPU, Intel Core i$7-12700$ processor, and $16.0$ GB RAM.

Next, we investigate the performance of the proposed scheme with and without compression.
{\small
\begin{table}[!t]
\caption{Training Details}
\label{tab_train}
\centering
\resizebox{\columnwidth}{!}{%
\begin{tabular}{|c|cccc|c}
\cline{1-5}
                                      & \multicolumn{2}{c|}{\text{Low Resolution (LR)}}                                            & \multicolumn{2}{c|}{\text{High Resolution (HR)}}                                   &  \\ \cline{1-5} 
\multirow{3}{*}{\text{Dataset}}     & \multicolumn{1}{c|}{\text{Training}}     & \multicolumn{1}{c|}{\text{Testing}}      & \multicolumn{1}{c|}{\text{Training}}             & \text{Testing}              &  \\ \cline{2-5}
                                      & \multicolumn{2}{c|}{CIFAR10}                                                            & \multicolumn{1}{c|}{DIV2K}                         & Kodak                         &  \\ \cline{2-5}
                                      & \multicolumn{2}{c|}{$32 \times 32$}                                                            & \multicolumn{1}{c|}{$2048 \times 2048$}                   & $768 \times 512$                     &  \\ \cline{1-5}
\multirow{3}{*}{\text{Swin Stages}} & \multicolumn{2}{c|}{\text{2 Stages}}                                                  & \multicolumn{2}{c|}{\text{4 Stages}}                                             &  \\ \cline{2-5}
                                      & \multicolumn{1}{c|}{\textbf{{[}$N_1$, $N_2${]}}} & \multicolumn{1}{c|}{\textbf{{[}$C_1$, $C_2${]}}} & \multicolumn{1}{c|}{\textbf{{[}$N_1$, $N_2$, $N_3$, $N_4${]}}} & \textbf{{[}$C_1$, $C_2$, $C_3$, $C_4${]}} &  \\ \cline{2-5}
                                      & \multicolumn{1}{c|}{{[}2, 4{]}}            & \multicolumn{1}{c|}{{[}$128$, $256${]}}        & \multicolumn{1}{c|}{{[}$2$, $2$, $6$, $2${]}}              & {[}$128$, $192$, $256$, $320${]}      &  \\ \cline{1-5}
\text{Batch Size}                   & \multicolumn{2}{c|}{$128$}                                                                & \multicolumn{2}{c|}{$16$}                                                            &  \\ \cline{1-5}
\end{tabular}
}
\end{table}
}
\subsection{Performance Evaluation Without Compression} \label{SubSec_Perf_W/O_C}
\subsubsection{CIFAR10 Dataset} \label{subsubsec_CIFAR10}

\begin{figure}[!ht]
\centering
\centerline{\includegraphics[width=\columnwidth]{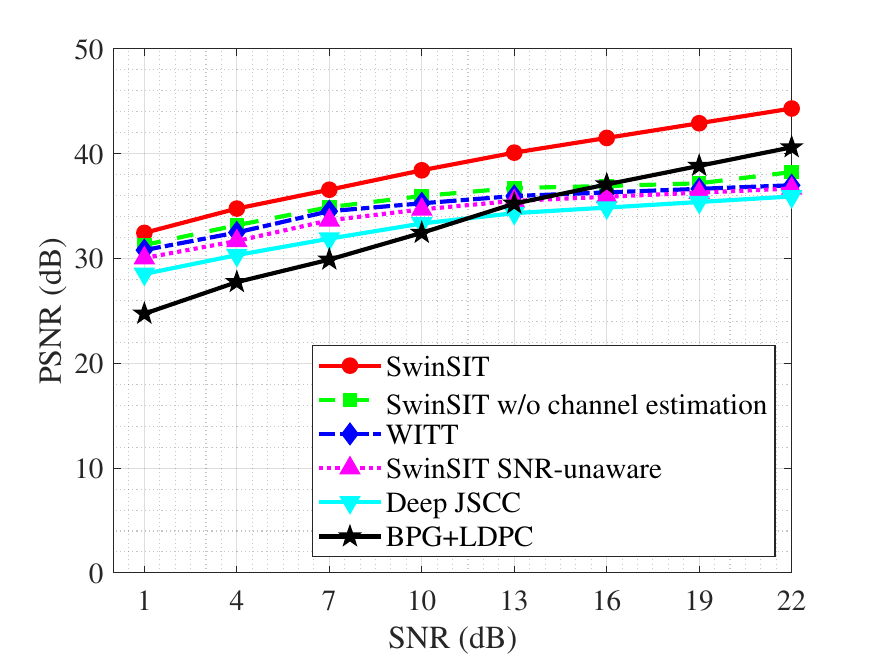}}
\caption{The effect of SNR on PSNR for R=$0.3$ with the CIFAR10 dataset.}
\label{fig_Res_CIFAR10_SNR_vs_PSNR}
\end{figure}

Figure \ref{fig_Res_CIFAR10_SNR_vs_PSNR} compares the PSNR performance of the SwinSIT model with the WITT model, the Deep JSCC model, and the BPG$+$LDPC scheme. The SwinSIT model exhibits superior PSNR gains over all the others. Also, the SwinSIT w/o channel estimation performs better than the others up to SNR of approximately $15$ dB. Beyond $15$ dB its performance degrades and falls below that of the BPG$+$LDPC scheme but still outperforms the rest. This ensures the importance of the proposed CEAC in achieving good performance.
\begin{figure}[!ht]
\centering
\centerline{\includegraphics[width=\columnwidth]{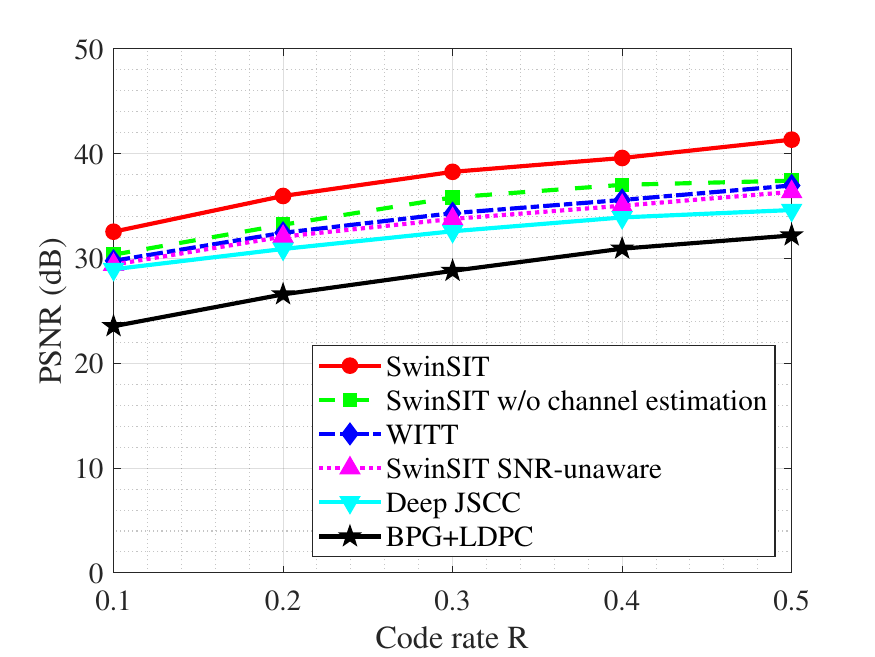}}
\caption{The impact of R on PSNR for the CIFAR10 dataset with SNR = $9$ dB.}
\label{fig_Res_CIFAR10_CBR_9dB}
\end{figure}
\begin{figure}[!ht]
\centering
\centerline{\includegraphics[width=\columnwidth]{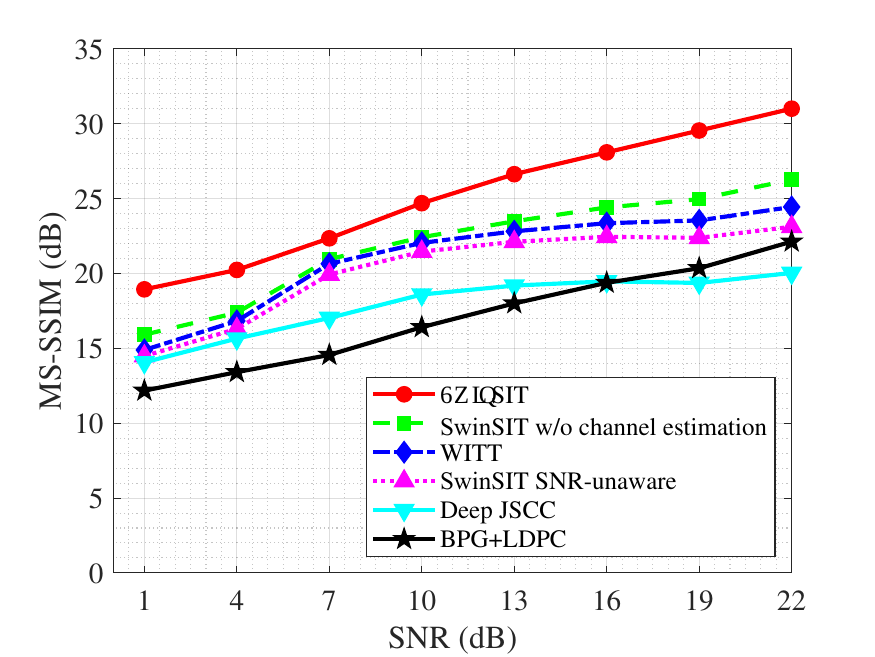}}
\caption{The MS-SSIM performance with respect to SNR for the CIFAR10 dataset.}
\label{fig_Res_CIFAR10_SSIM}
\end{figure}

Figure \ref{fig_Res_CIFAR10_CBR_9dB} shows the effects of the code rate $R$ on PSNR at SNR of $9$ dB. The results show that the BPG$+$LDPC scheme provides lower PSNR, while CNN-based approaches achieve better performance. However, the proposed scheme, even w/o channel estimation, outperforms all the others with respect to PSNR. Besides, the results show that the SwinSIT SNR-unaware still competes with the WITT model and outperforms the Deep JSCC model.

In Fig. \ref{fig_Res_CIFAR10_SSIM}, the MS-SSIM performance against the SNR is illustrated. Notably, the MS-SSIM offers a more human-centric evaluation by considering structural and textural elements aligning better with how we perceive images. Since the SwinSIT SNR-unaware has no SNR-aware capabilities, it has lower MS-SSIM values than the SwinSIT SNR-aware and WITT models. It can be seen that the SwinSIT still provides the best MS-SSIM performance among the others, whereas the BPG+LDPC scheme yields the lowest MS-SSIM, especially for low and moderate SNRs.

\subsubsection{DIV2K-Kodak Datasets} \label{subsubsec_DIV2K}
\begin{figure}[!ht]
\centering
\centerline{\includegraphics[width=\columnwidth]{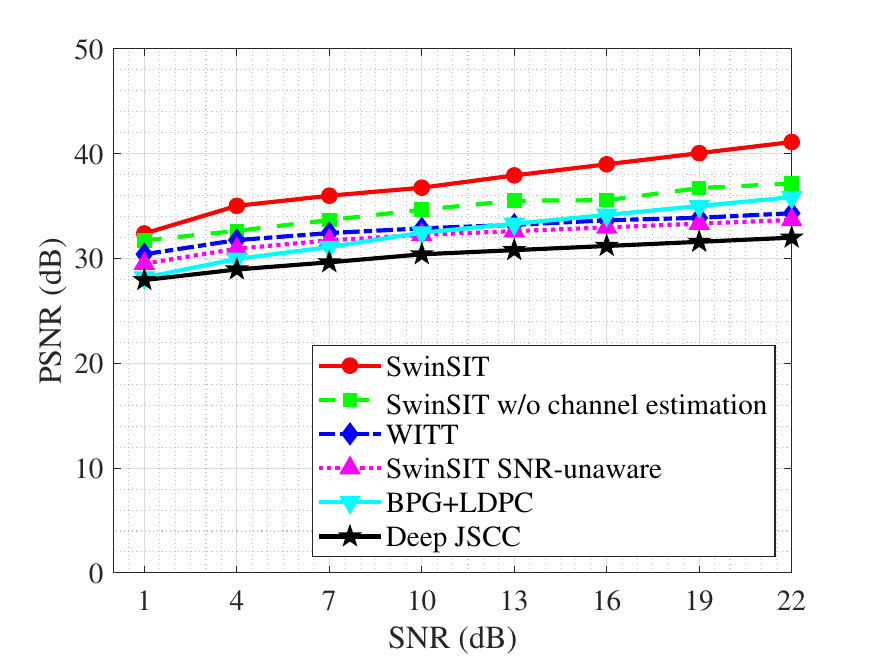}}
\caption{The effect of SNR on PSNR for R=$0.0625$ with the DIV2K-Kodak datasets.}
\label{fig_Res_DIV2K_SNR_vs_PSNR}
\end{figure}

In Fig. \ref{fig_Res_DIV2K_SNR_vs_PSNR}, the PSNR results are investigated over different SNRs. The SwinSIT leads the others as in LR results, reaffirming the superior performance in HR images. Compared to Fig. \ref{fig_Res_CIFAR10_SNR_vs_PSNR}, there is a bigger gap between the curves of SwinSIT and the SwinSIT w/o channel estimation. Also, the CNN-based Deep JSCC scheme performs poorly when handling HR images, resulting in lower performance compared to the BPG$+$LDPC scheme. The performance of the BPG$+$LDPC scheme improves slightly compared to the WITT model as SNR increases.

\begin{figure}[!ht]
\centering
\centerline{\includegraphics[width=\columnwidth]{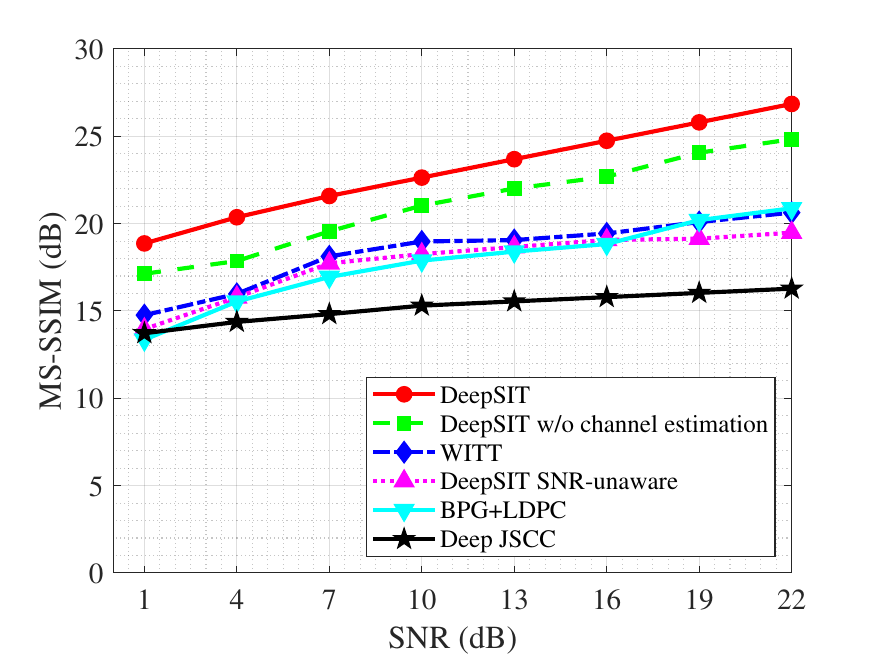}}
\caption{The MS-SSIM with respect to SNR for the DIV2K-Kodak datasets.}
\label{fig_Res_DIV2K_SSIM}
\end{figure}

In Fig. \ref{fig_Res_DIV2K_SSIM}, the MS-SSIM performance is inspected with respect to SNR. It can be checked that the SwinSIT still outperforms the others with respect to the MS-SSIM. Also, it is clear that the CNN-based Deep JSCC model has the worst performance. We can conclude that transformer-based models such as our proposed scheme and the WITT model show superior results than other CNN-based models. However, for high SNR values, the BPG$+$LDPC scheme shows better performance than the WITT and SwinSIT SNR-unaware models.

\subsection{Performance Evaluation With Compression} \label{SubSec_Perf_W_C}

Here, we introduce a compressed version of the SwinSIT model, achieved through the joint pruning and quantization process outlined in Algorithm~\ref{Alg_P&Q}. Specifically, we apply a 60\% pruning rate and quantize both weights and activations from high-precision FP$32$ to low-precision INT$8$ optimizing the model for resource-constrained IoT devices by reducing computational complexity and memory requirements.

Table~\ref{tab_P&Q} presents a comparison of different SwinSIT variants: the original model, the pruned model (60\% pruning), and the fully compressed model (60\% pruning with FP$32$ to INT$8$ quantization). Pruning reduces the number of parameters, shrinking the model size from $53.4$ MB to $21.3$ MB. Further quantization significantly compresses it to $5.3$ MB making it more suitable for IoT applications while maintaining competitive performance, as demonstrated in the subsequent results.
\begin{table}[!ht]
\caption{Comparing SWinSIT Model Before and After Compression.}
\label{tab_P&Q}
\centering
\scriptsize
\resizebox{0.95\columnwidth}{!}{%
\begin{tabular}{|c|c|c|c|}
\hline
\multirow{2}{*}{} & \multicolumn{1}{c|}{Original} & \multicolumn{1}{c|}{After Pruning} & \multicolumn{1}{c|}{After Pruning } \\ 
 & & & and Quantization \\ \hline
Total Parameters & 13,679,488 & 8,207,693 & 5,471,795 \\ \hline
Model Size       & 53.4 MB    & 21.3 MB    & 5.3 MB    \\ \hline
Pruned Parameters & N/A        & 8,207,693  & 8,207,693       \\ \hline
Remaining Parameters & N/A      & 5,471,795  & 5,471,795       \\ \hline
\end{tabular}}
\end{table}

\begin{figure}[!ht]
\centering
\centerline{\includegraphics[width=\columnwidth]{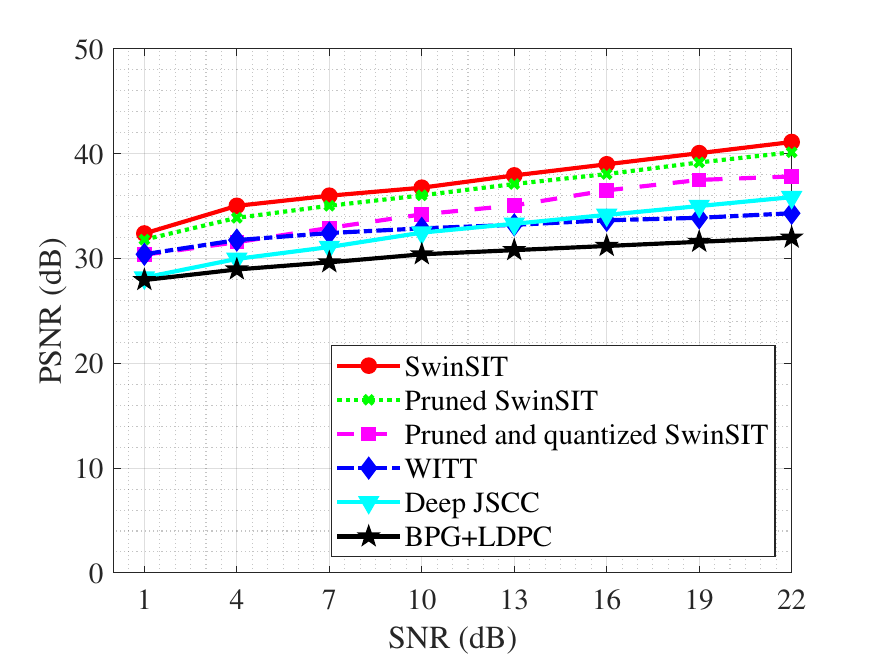}}
\caption{PSNR performance as a function of SNR for the compressed SwinSIT model on the DIV2K-Kodak dataset.}
\label{fig_DIV2K_SNR_PSNR_PQ}
\end{figure}

Figures \ref{fig_DIV2K_SNR_PSNR_PQ} illustrates the performance of the compressed SwinSIT model on the DIV2K-Kodak dataset. The figure compares the PSNR performance across different SNR values for SwinSIT, pruned SwinSIT, and pruned-quantized SwinSIT against other benchmarks. The results demonstrate that pruning and quantization effectively reduce model complexity while maintaining competitive performance with a manageable degradation. Specifically, the pruned model continues to outperform other approaches while requiring fewer resources. Although the pruned-quantized model exhibits slightly lower performance than WITT, it still achieves favorable results with significantly reduced complexity. Depending on IoT network constraints, a tradeoff between model size and image reconstruction quality can be optimized, leveraging the flexibility of the proposed compression techniques.
\section{Conclusion} \label{Sec_Conc}
In this paper, we have studied the problem of designing an SC system for image transmission based on a JSCC paradigm for the Rayleigh fading channel. The SwinSIT model has been proposed to exploit the hierarchical global semantic extraction of the Swin transformer for constructing both the encoder and decoder. To deal with the SNR variations, a CNN-based SNR-aware module has been presented inspired by the SE which performed a double-phase enhancement for the extracted semantic map and its noisy version. To accurately estimate the channel fading, a new DnCNN-based CEAC has been adopted based on CNN networks. To enhance the practicality of SwinSIT in resource-constrained IoT environments, a joint pruning and quantization scheme has been proposed, significantly reducing model complexity while maintaining high reconstruction quality. The simulation results have demonstrated that the proposed SwinSIT model outperformes the conventional models in terms of PSNR and MS-SSIM. Furthermore, the compressed version of the proposed model reduces its size while preserving favorable PSNR performance demonstrating its efficiency and robustness for SC-based image transmission.

\bibliographystyle{IEEEtran}
\bibliography{Main}

 




\vfill

\end{document}